\documentclass[%
 reprint,
superscriptaddress,
showkeys,amsmath,amssymb,aps,]{revtex4-2}

\usepackage{graphicx}
\usepackage{dcolumn}
\usepackage{bm}

\begin{document}

\preprint{APS/123-QED}

\title{$\mathcal{PT}$-Symmetric Quantum Discrimination of Three States}

\author{Yaroslav Balytskyi}
\email{ybalytsk@uccs.edu}
\affiliation{Department of Physics and Energy Science, University of Colorado, Colorado Springs, Colorado, 80918, USA}

\author{Manohar Raavi}
\email{mraavi@uccs.edu}
\affiliation{Department of Computer Science, University of Colorado, Colorado Springs, Colorado, 80918, USA}

\author{Anatoliy Pinchuk}
\email{apinchuk@uccs.edu}
\affiliation{Department of Physics and Energy Science, University of Colorado, Colorado Springs, Colorado, 80918, USA}

\author{Sang-Yoon Chang}
\email{schang2@uccs.edu}
\affiliation{Department of Computer Science, University of Colorado, Colorado Springs, Colorado, 80918, USA}

\date{\today}

\begin{abstract}

If the system is known to be in one of two non-orthogonal quantum states, $|\psi_1\rangle$ or $|\psi_2\rangle$, it is not possible to discriminate them by a single measurement due to the unitarity constraint. In a regular Hermitian quantum mechanics, the successful discrimination is possible to perform with the probability $p < 1$, while in $\mathcal{PT}$-symmetric quantum mechanics a \textit{simulated single-measurement} quantum state discrimination with the success rate $p$ can be done. We extend the $\mathcal{PT}$-symmetric quantum state discrimination approach for the case of three pure quantum states, $|\psi_1\rangle$, $|\psi_2\rangle$ and $|\psi_3\rangle$ without any additional restrictions on the geometry and symmetry possession of these states. We discuss the relation of our approach with the recent implementation of $\mathcal{PT}$ symmetry on the IBM quantum processor.

\end{abstract}


\maketitle
\section{Introduction}

Quantum state discrimination is a central problem in many applications in quantum computing, communications, cryptography and information processing as well as in the research on the foundations of the quantum theory; see a review article ~\cite{bae2015quantum}. It is formulated as follows. Initially, Alice and Bob agree on a set of possible signal states which are "letters of the alphabet" and the prior probabilities of each state to occur are assumed to be known to both parties. Alice encodes the message by the set of these states and Bob's task is to find the state of the system by one or more measurements in the most optimal way. This goal can be achieved by the POVM measurements which are a set of positive operators $\{ F_m\}$ acting on the Hilbert space and in the sum being the identity matrix $\sum_{m}F_m = I$. If the quantum system is initialized in the state $|\chi\rangle$, the probability of obtaining the result $m$ is $\langle\chi|F_m|\chi\rangle$, and the state of the system after the measurement is:

$$
|\chi\rangle\rightarrow\frac{\sqrt{F_{m}}|\chi\rangle}{\sqrt{\langle\chi|F_m|\chi\rangle}}
$$

For the case of $N = 2$ states, a simple classical analog is the following. Bob (experimenter) is given a priori knowledge that the coin can be either: fair ($50 \% /50\%$) or biased ($51\% / 49\%$). Tossing the coin many times, he needs to determine which option is true. There is a number of applications of the quantum state discrimination problem. For example, in quantum information processing, quantum state discrimination plays a crucial role in deriving the no-go theorem for an interpretation of quantum states ~\cite{pusey2012}. Additionally, it is also needed for the search for a dimension witness of quantum systems ~\cite{Brunner2013}, and as an operational interpretation of conditional mutual entropy ~\cite{Koenig2009}. In quantum cryptography, the security of the quantum key distribution is based on the difficulty of a quantum state discrimination ~\cite{BB84}, and on the no-cloning theorem ~\cite{nocloning}. In quantum computation it is established that an unstructured database search can be mapped to the problem of discriminating the quantum states that are exponentially close to each other ~\cite{Abrams}.

On the other hand, $\mathcal{PT}$-symmetric quantum mechanics has emerged as an extension of the usual Hermitian one. In this approach, $\mathcal{PT}$ symmetry replaces the condition of Hermiticity of the Hamiltonian, $\mathcal{H} = \mathcal{H}^{\dagger}$. The Hamiltonian $\mathcal{H}$ is told to possess $\mathcal{PT}$ symmetry if it satisfies the requirement $\mathcal{H} = \mathcal{H}^{\mathcal{PT}}$.
The parity operator $\mathcal{P}$  changes the sign of quantum-mechanical
coordinate $\hat{x}$ and the momentum $\hat{p}$ operators: 

$$
\mathcal{P}\hat{x}\mathcal{P} = - \hat{x}; \ \ \mathcal{P}\hat{p}\mathcal{P} = - \hat{p}
$$

and in the $2\times2$ case is given by: 

$$
\mathcal{P} =
\begin{pmatrix}
0 & 1 \\
1 & 0
\end{pmatrix}  
$$

The time-reversal operator $\mathcal{T}$ operator leaves $\hat{x}$ invariant but changes the sign of $\hat{p}$ and the imaginary unit $i$:

$$
\mathcal{T}\hat{x}\mathcal{T} = \hat{x}; \ \ \mathcal{T}\hat{p}\mathcal{T} = - \hat{p}; \ \  \mathcal{T}i\mathcal{T} = - i
$$

Remarkably, while in the Hermitian case the inner product is fixed, $\mathcal{PT}$-symmetric Hamiltonian \textit{determines} an inner product giving an extra degree of freedom. This additional degree of freedom can be used to speed up the quantum evolution. This effect was theoretically predicted~\cite{bender2007faster}, and the speed up was observed experimentally~\cite{zheng2013observation}. 

Additionally, the application of this degree of freedom enables to discriminate two non-orthogonal states, \textit{in principle}, by a single measurement~\cite{bender2013pt}. However, in such an approach, the initial Hilbert space of the system in which the state vectors are defined is changed to a new one spanned on the eigenvectors of the $\mathcal{PT}$-symmetric Hamiltonian, and this change is carried out by a similarity transformation which introduces a probability that the measurement produces a null result. As a result, the evolution of the projected wave function resulting in a perfect state discrimination is described by the $\mathcal{PT}$-symmetric Hamiltonian. However, the norm of the corresponding projected component of the wave function is less than one, and an inconclusive result is obtained with probability greater than zero. Therefore, while, \textit{in principle}, to distinguish $N = 2$ states we need only a single measurement, even in the ideal noiseless case application of more than one measurement may be required since some of them may be inconclusive. In this sense $\mathcal{PT}$-symmetric quantum discrimination approach is similar to an unambiguous state discrimination~\cite{Ivanovic}.

A general $\mathcal{PT}$-symmetric Hamiltonian has the following form ~\cite{bender2013pt}: 
\begin{equation}\label{Hamiltonian}   
\mathcal{H} = \mathcal{H}^{\mathcal{PT}} = 
\begin{pmatrix}
r e^{i\theta} & s \\
s & r e^{-i\theta}
\end{pmatrix}
\end{equation}

with $r$, $s$ and $\theta$ being real parameters. The additional degree of freedom provided by $\mathcal{PT}$ symmetry is represented by $\alpha$ parameter which is defined by the parameters of $\mathcal{PT}$-symmetric Hamiltonian, Eqn.(\ref{Hamiltonian}), $\sin\left(\alpha\right) = \frac{r}{s}\sin\left(\theta\right)$. An inner product is defined by the $\mathcal{C}$ operator which commutes with $\mathcal{H}$ and $\mathcal{PT}$:
$$
\left[\mathcal{C}, \mathcal{H}\right] = 0, \ \left[\mathcal{C}, \mathcal{PT}\right] = 0
$$

and which is absent in a regular Hermitian case:

$$
\mathcal{C} =\frac{1}{\cos\left(\alpha\right)}
\begin{pmatrix}
i \sin\left( \alpha \right) & 1 \\
1 & -i \sin\left( \alpha \right)
\end{pmatrix}  
$$

The bra-vector of the state $|\mu\rangle$ is given by: 

\begin{equation}
  \left(\langle\mu|\right)_{\mathcal{CPT}} = \left(\mathcal{CPT}|\mu\rangle\right)^T
  \label{left}
\end{equation}

and the corresponding $\mathcal{CPT}$ scalar product of two vectors $|\mu\rangle$ and $|\nu\rangle$:

$$
\langle\mu|\nu\rangle = \left(\mathcal{CPT} \mu\right)^T\cdot\nu
\label{CPT}
$$

where the superscript $T$ means the matrix transposition. For the number of reference states $N = 2$ two alternative approaches are possible~\cite{bender2013pt}:
\begin{itemize}
\item \textit{Solution 1: Adjusting $\mathcal{PT}$-symmetric Hamiltonian in such a way that its $\mathcal{CPT}$ inner product interprets $|\psi_1\rangle$ and $|\psi_2\rangle$ as being orthogonal.}
\item \textit{Solution 2:
Using $\mathcal{PT}$-symmetric Hamiltonian to evolve the states $|\psi_1\rangle$ and $|\psi_2\rangle$  into orthogonal under the Hermitian scalar product.}
\end{itemize}

Moreover, recently the $\mathcal{PT}$ symmetric quantum state discrimination corresponding to \textit{Solution 2} was investigated using the emergent technology of superconducting quantum processors~\cite{IBM}, substantial progress on which was achieved by IBM, by employing the dilation method involving an ancillary qubit~\cite{PTIBM}.

The goal of our paper is to extend the $\mathcal{PT}$-symmetric quantum state discrimination approach for the case of $N = 3$ pure quantum states. In general, for the case when $N > 2$, optimal state discrimination is solved only under the conditions of possessing particular symmetries such as "geometrically uniform" states~\cite{Uniform} and mirror symmetric states~\cite{Mirror}. The usage of an additional degree of freedom provided by $\mathcal{PT}$ symmetry allows to relax an assumption about the possession of certain symmetries of quantum states. In our approach, we use the combination of the application of the gates and both $\textit{Solution 1}$ and $\textit{Solution 2}$. Under the same assumptions used for the two-state $\mathcal{PT}$-symmetric discrimination, we extend this approach to $N = 3$ case and show that \textit{in principle} it is possible to find the state of the system with at most two identical prepared samples for the measurements.

\section{Extension to $N = 3$ quantum states}
Our approach involves three steps:

\begin{itemize}
    \item \textit{Step 1}: With the use of \textit{Solution 2},  adjust the first two vectors $|\psi_1\rangle$ and $|\psi_2\rangle$ on the opposite sides of the Bloch sphere, i.e. make them orthogonal in a sense of the Hermitian scalar product.
    \item \textit{Step 2}: By applying the gates, adjust the first two states in the following way:
    $$
|\psi_{1,2}\rangle \rightarrow \frac{1}{\sqrt{2}}\begin{pmatrix}
1  \\
\pm i
\end{pmatrix}
$$
and according to Eqn.~\ref{left}:
    $$
\left(\langle\psi_{1,2}|\right)\mathcal{CPT} \rightarrow \frac{1 \pm \sin\left(\alpha\right)}{\sqrt{2}}\begin{pmatrix}
1  \\
\mp i
\end{pmatrix}
$$
As a result, the first two states are orthogonal under the $\mathcal{CPT}$ inner product for an arbitrary value of $\mathcal{\alpha}$ which we are free to adjust.
\item \textit{Step 3}: Vary the $\alpha$ parameter in order to adjust the angle between $|\psi_{1,2}\rangle$ and $|\psi_3\rangle$, and apply the $\mathcal{CPT}$ measurement to eliminate one of the states. 
\end{itemize}

We start from three completely arbitrary states without any assumption on possessing any symmetry: 

$$ 
|\psi_i\rangle = \begin{pmatrix}
\cos\left(\frac{\theta_i}{2}\right)  \\
e^{i\phi_i}\sin\left(\frac{\theta_i}{2}\right) 
\end{pmatrix}; i \in [1, 3]
$$

where $\theta_i$ are the parallels and $\phi_i$ are the meridians of the positions of the $|\psi_i\rangle$ vectors on the Bloch sphere. We assume these three states to have the prior probabilities to be $p_i$. Without loss of generality, we assume the first state $|\psi_1\rangle$ to have to highest prior probability: $p = \underset{i \in [1, 3]}{max} \{p_i\}$ and the aim of the first measurement is to eliminate the state with the highest prior probability first. 

We use the following gates to adjust our states in a convenient positions for the Hamiltonian evolution by \textit{Solution 2}:

$$
R_1 = 
\begin{pmatrix}
\cos\left(\frac{\theta_1}{2}\right) & \sin\left(\frac{\theta_1}{2}\right)e^{-i\phi_1}\\
 -\sin\left(\frac{\theta_1}{2}\right)e^{i\phi_1} & \cos\left(\frac{\theta_1}{2}\right) 
\end{pmatrix}
$$
$$
R_2 = 
\begin{pmatrix}
1 & 0\\
0& -i e^{- i\lambda - i\phi_2}
\end{pmatrix}
$$

$$
R_3 = 
\begin{pmatrix}
\cos\left(\frac{\pi - 2\sigma}{4}\right) & -i\sin\left(\frac{\pi - 2\sigma}{4}\right)\\
-i\sin\left(\frac{\pi - 2\sigma}{4}\right)& \cos\left(\frac{\pi - 2\sigma}{4}\right)
\end{pmatrix}
$$

where $\sigma$, and $\lambda$ parameters are given in the Eqn.\ref{PrepareHamiltonian1} and Eqn.\ref{PrepareHamiltonian2}. These gates transform our states as: 
$$
|\psi_i\rangle \rightarrow R_3R_2R_1|\psi_i\rangle,  k \in [1, 3]
$$

After their application, the first two our states take the form convenient for the subsequent Hamiltonian evolution:

$$
|\psi_{1,2}\rangle \rightarrow \begin{pmatrix}
\cos\left(\frac{\pi \mp 2\sigma}{4}\right)  \\
-i\sin\left(\frac{\pi \mp 2\sigma}{4}\right)
\end{pmatrix}
$$
The third state takes the following form:
$$
|\psi_3\rangle \rightarrow \begin{pmatrix}
\cos\left(\frac{\mu}{2}\right)  \\
e^{i\nu}\sin\left(\frac{\mu}{2}\right)
\end{pmatrix}
$$
with the $\mu$ and $\nu$ parameters given by an Eqn.\ref{PrepareHamiltonian3}.

\begin{widetext}

\begin{equation}
\begin{aligned}\label{PrepareHamiltonian1}
  \cos\left(\sigma\right) = \sqrt{\frac{1 +\cos(\text{$\theta_1$}) \cos (\text{$\theta_2$}) + \sin (\text{$\theta_1$})\sin (\text{$\theta_2$}) \cos(\text{$\phi_1$}-\text{$\phi_2$})}{2}}
  \end{aligned}
\end{equation}

\begin{equation}
  \begin{aligned}
     \label{PrepareHamiltonian2}\lambda  =\arctan\left(\frac{\sin\left(\frac{\theta_1}{2}\right)\cos\left(\frac{\theta_2}{2}\right)\sin\left(\phi_2 - \phi_1\right)}{\cos\left(\frac{\theta_1}{2}\right)\sin\left(\frac{\theta_2}{2}\right) - \sin\left(\frac{\theta_1}{2}\right)\cos\left(\frac{\theta_2}{2}\right)\cos\left(\phi_2 - \phi_1\right)}\right) - \\- \arctan\left(\frac{\sin\left(\frac{\theta_1}{2}\right)\sin\left(\frac{\theta_2}{2}\right)\sin\left(\phi_2 - \phi_1\right)}{\cos\left(\frac{\theta_1}{2}\right)\cos\left(\frac{\theta_2}{2}\right) + \sin\left(\frac{\theta_1}{2}\right)\sin\left(\frac{\theta_2}{2}\right)\cos\left(\phi_2 - \phi_1\right)}\right) 
\end{aligned}
\end{equation}
\end{widetext}

\begin{widetext}
\begin{equation}
\begin{aligned}\label{PrepareHamiltonian3}
          \beta = \cos\left(\frac{\theta_1}{2}\right)\cos\left(\frac{\theta_3}{2}\right)\cos\left(\frac{\pi - 2\sigma}{4}\right)\left(1 + \tan\left(\frac{\theta_1}{2}\right)\tan\left(\frac{\pi - 2\sigma}{4}\right)e^{i\phi_1 - i\phi_2 - i\lambda}\right) + \\
     + \sin\left(\frac{\theta_1}{2}\right)\sin\left(\frac{\theta_3}{2}\right)\cos\left(\frac{\pi - 2\sigma}{4}\right)e^{i\phi_3 - i\phi_1}\left(1 - \cot\left(\frac{\theta_1}{2}\right)\tan\left(\frac{\pi - 2\sigma}{4}\right)e^{i\phi_1 - i\phi_2 - i\lambda}\right)
     \\     \gamma = i\cos\left(\frac{\theta_1}{2}\right)\cos\left(\frac{\theta_3}{2}\right)\sin\left(\frac{\pi - 2\sigma}{4}\right)\left(\tan\left(\frac{\theta_1}{2}\right)\cot\left(\frac{\pi - 2\sigma}{4}\right)e^{i\phi_1 - i\phi_2 - i\lambda} - 1\right) - \\
     - i\sin\left(\frac{\theta_1}{2}\right)\sin\left(\frac{\theta_3}{2}\right)\sin\left(\frac{\pi - 2\sigma}{4}\right)e^{i\phi_3 - i\phi_1}\left(1 + \cot\left(\frac{\theta_1}{2}\right)\cot\left(\frac{\pi - 2\sigma}{4}\right)e^{i\phi_1 - i\phi_2 - i\lambda}\right)
     \\
    \cos\left(\frac{\mu}{2}\right) = \sqrt{\left(Re\left(\beta\right)\right)^2 + \left(Im\left(\beta\right)\right)^2} = |\beta|; 
     \ \nu = \arctan\left(\frac{Im\left(\gamma\right)}{Re\left(\gamma\right)}\right) - \arctan\left(\frac{Im\left(\beta\right)}{Re\left(\beta\right)}\right)
\end{aligned}
\end{equation}
\end{widetext}

Then we apply the Hamiltonian evolution:
$$ 
e^{-i\mathcal{H}t} = \frac{e^{-i r\cos\left(\theta\right)t}}{\cos\left(\alpha\right)}
\begin{pmatrix}
\cos\left(\omega t - \alpha\right) & -i\sin\left(\omega t \right) \\
-i\sin\left(\omega t \right) & \cos\left(\omega t + \alpha\right)
\end{pmatrix}  
$$

where $\omega = \sqrt{s^2 - r^2\sin^2\theta}$. Note that in the case when $\alpha = 0$, $e^{+i\mathcal{H}t}e^{-i\mathcal{H}t} = \hat{1}$. Therefore, in order to make such an evolution successful, we choose the value of parameter $\alpha$ to be nonzero.  We will specify its precise value further in the text.

After the time $\tau$ given by an equation:

$$
\sin^2\left(\omega \tau\right) = \frac{\cos^2\alpha \cos\sigma}{2\sin\alpha - 2\sin^2\alpha\cos\sigma}
$$

the first two states become orthogonal in a sense of the Hermitian scalar product:
$$(\langle\psi_1|\psi_2\rangle)_{Hermitian} = 0$$

After the Hamiltonian evolution, the first two states are transformed to the following form:
$$
|\psi_1\rangle \rightarrow \begin{pmatrix}
\cos\left(\frac{\delta}{2}\right) \\
-i\sin\left(\frac{\delta}{2}\right)
\end{pmatrix};  \ |\psi_2\rangle \rightarrow \begin{pmatrix}
\sin\left(\frac{\delta}{2}\right)  \\
i\cos\left(\frac{\delta}{2}\right)
\end{pmatrix}
$$

where the $\delta$ parameter is given by Eqn.\ref{Evolved}.

Next, we apply the following two gates with the $\chi$ parameter given by Eqn.\ref{AfterEvolution1}:
$$
R_4 = 
\begin{pmatrix}
\cos\left(\frac{\delta}{2}\right) & i\sin\left(\frac{\delta}{2}\right)\\
i\sin\left(\frac{\delta}{2}\right) & \cos\left(\frac{\delta}{2}\right)
\end{pmatrix}
$$

$$
R_5 = 
\begin{pmatrix}
1 & 0\\
0 & ie^{-i\chi}
\end{pmatrix}
$$

After application of these gates, the first two of our states are on the North and South poles of the Bloch sphere while third one is specified by the parameter $\xi$ provided by Eqn.\ref{AfterEvolution1}:
$$
|\psi_1\rangle \rightarrow \begin{pmatrix}
1 \\
0
\end{pmatrix};  \ |\psi_2\rangle \rightarrow \begin{pmatrix}
0  \\
1
\end{pmatrix};
|\psi_3\rangle \rightarrow \begin{pmatrix}
\cos\left(\frac{\xi}{2}\right) \\
i\sin\left(\frac{\xi}{2}\right)
\end{pmatrix}
$$

\begin{widetext}{}
\begin{equation}
  \begin{aligned}\label{Evolved}
     \cos\left(\frac{\delta}{2}\right) = \frac{\cos\left(\omega\tau - \alpha\right)\cos\left(\frac{\pi - 2\sigma}{4}\right)  - \sin\left(\omega\tau\right)\sin\left(\frac{\pi - 2\sigma}{4}\right)}{\sqrt{1 - \cos\left(2\omega\tau\right)\sin^2\left(\alpha\right) +  2\sin\left(\omega\tau\right)\sin\left(\alpha\right) \left(\cos\left(\omega\tau\right)\cos\left(\alpha\right)\sin\left(\sigma\right) - \sin\left(\omega\tau\right)\cos\left(\sigma\right) \right)}}
  \end{aligned}
  \end{equation}
\end{widetext}

\begin{widetext}{}
\begin{equation}
  \begin{aligned}
     \label{AfterEvolution1}\kappa = \cos\left(\frac{\mu}{2}\right)\left( \cos\left(\omega\tau - \alpha\right)\cos\left(\frac{\delta}{2}\right) + \sin\left(\omega\tau \right)\sin\left(\frac{\delta}{2}\right) \right) + i e^{i\nu} \sin\left(\frac{\mu}{2}\right)\left( \cos\left(\omega\tau + \alpha\right)\sin\left(\frac{\delta}{2}\right) - \sin\left(\omega\tau \right)\cos\left(\frac{\delta}{2}\right) \right) 
\\
\zeta = i\cos\left(\frac{\mu}{2}\right)\left( \cos\left(\omega\tau - \alpha\right)\sin\left(\frac{\delta}{2}\right) - \sin\left(\omega\tau \right)\cos\left(\frac{\delta}{2}\right) \right) + e^{i\nu} \sin\left(\frac{\mu}{2}\right)\left( \cos\left(\omega\tau + \alpha\right)\cos\left(\frac{\delta}{2}\right) + \sin\left(\omega\tau \right)\sin\left(\frac{\delta}{2}\right) \right) 
     \\ 
\cos\left(\frac{\xi}{2}\right)  = \frac{|\kappa|}{\sqrt{|\kappa|^2 + |\zeta|^2}}; \ \chi = \arctan\left(\frac{Im\left(\zeta\right)}{Re\left(\zeta\right)}\right) - \arctan\left(\frac{Im\left(\kappa\right)}{Re\left(\kappa\right)}\right)
  \end{aligned}
  \end{equation}
\end{widetext}
Additionally, we apply the following gate:
$$
R_6 = \frac{1}{\sqrt{2}}
\begin{pmatrix}
1 & i\\
i & 1
\end{pmatrix}
$$

which puts our states in a convenient positions for the $\mathcal{CPT}$ measurements and $\rho = \xi + \frac{\pi}{2}$:
$$
|\psi_1\rangle \rightarrow \frac{1}{\sqrt{2}}\begin{pmatrix}
1  \\
i
\end{pmatrix};  \ |\psi_2\rangle \rightarrow \frac{1}{\sqrt{2}}\begin{pmatrix}
1  \\
- i
\end{pmatrix};
\ |\psi_3\rangle \rightarrow \begin{pmatrix}
\cos\left(\frac{\rho}{2}\right)  \\
i\sin\left(\frac{\rho}{2}\right)
\end{pmatrix}
$$

Note that the $\mathcal{CPT}$ scalar product of the first two states is zero for an arbitrary value of the $\alpha$ parameter: 

$$\left(\langle\psi_1|\psi_2\rangle \right)_{\mathcal{CPT}} = 0
$$

since 

$$
\left(\langle\psi_1|\right)_{\mathcal{CPT}} = \frac{\left(1 + \sin\left(\alpha\right)\right)}{\sqrt{2}\cos\left(\alpha\right)}
\begin{pmatrix}
1 \\
-i
\end{pmatrix}^T
$$

This makes the cosine of the angle between the first two states zero and allows us to vary the value of $\alpha$ for adjusting the relative angles  between the the first/second and the third state, $\kappa_{13}$ and $\kappa_{23}$:

$$
\cos^2\left(\kappa_{12}\right) = 0 
$$

The resulting cosines squared of the angles between these three states $\kappa_{13}$ and $\kappa_{23}$ are given by:
$$
\cos^2\left(\kappa_{13}\right) = \frac{\left(1+ \sin\left(\alpha\right)\right)\left(1+ \sin\left(\rho\right)\right)}{2\left(1 + \sin\left(\alpha\right)\sin\left(\rho\right)\right)}
$$

$$
\cos^2\left(\kappa_{23}\right) = \frac{\left(1- \sin\left(\alpha\right)\right)\left(1- \sin\left(\rho\right)\right)}{2\left(1 + \sin\left(\alpha\right)\sin\left(\rho\right)\right)}
$$

\begin{figure}[h!]
  \centering
  \includegraphics[width=8cm]{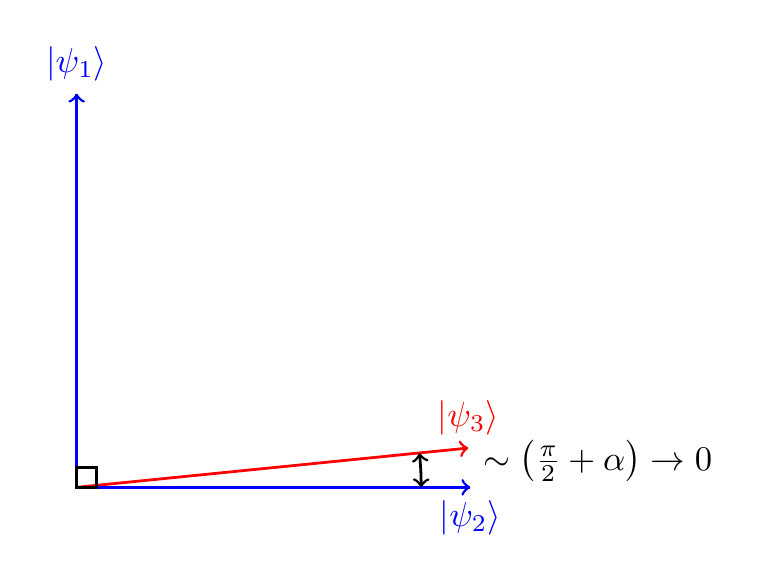}
  \caption{The geometry of the three states when $
  \kappa_{23}\sim \left(\frac{\pi}{2} + \alpha\right)\rightarrow0$.}
  \label{Angles_1}
\end{figure}

\begin{figure}[h!]
  \centering
  \includegraphics[width=8cm]{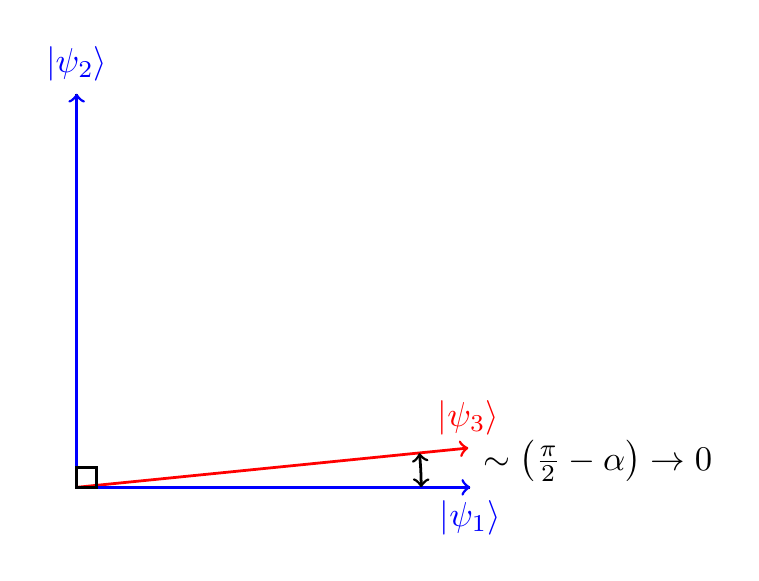}
  \caption{The geometry of the three states when $
  \kappa_{13}\sim \left(\frac{\pi}{2} - \alpha\right)\rightarrow0$.}
  \label{Angles_2}
\end{figure}

In the limit when $\alpha\rightarrow-\frac{\pi}{2}$ and $\kappa_{23}\rightarrow0$, the geometry of the states is shown on Fig.\ref{Angles_1}, and for the case when $\alpha\rightarrow\frac{\pi}{2}$ and $\kappa_{13}\rightarrow0$ on the Fig.\ref{Angles_2} correspondingly.

Adjusting our Hamiltonian (and simultaneously the $\mathcal{CPT}$ scalar product) in such as way that $\alpha\rightarrow-\frac{\pi}{2}$ which corresponds to Fig.\ref{Angles_1}, we are able to eliminate the first state $|\psi_1\rangle$ which has the highest prior probability by a single measurement.

The $\mathcal{CPT}$ projection operators are:
$$
\hat{P_{1}} = \left(\frac{|\psi_1\rangle\langle\psi_1|}{\langle\psi_1|\psi_1\rangle}\right)_{\mathcal{CPT}}=  \frac{1}{{2}}
\begin{pmatrix}
1 & -i\\
i & 1
\end{pmatrix}
$$

$$
\hat{P_{2}} = \left(\frac{|\psi_2\rangle\langle\psi_2|}{\langle\psi_2|\psi_2\rangle}\right)_{\mathcal{CPT}}=  \frac{1}{{2}}
\begin{pmatrix}
1 & i\\
-i & 1
\end{pmatrix}
$$

These operators are the $\mathcal{CPT}$ observables since:
$$
\left[\mathcal{CPT}, \hat{P}_{1, 2}\right] = 0
$$
Taking the limit $\alpha\rightarrow-\frac{\pi}{2}$ and applying the measurement on the first sample $
\hat{\mathcal{M}}|\psi_{Sample_1}\rangle
$ where:
$$
\hat{\mathcal{M}} = \hat{P_1} - \hat{P_2}
$$
we can determine whether the state we are looking for is $|\psi_1\rangle$ or one of $|\psi_2\rangle$ and $|\psi_3\rangle$:
$$
\begin{cases}
\mathcal{M} = 1 \Rightarrow |\psi_1\rangle \\
\mathcal{M} = -1 \Rightarrow |\psi_2\rangle \ or \ |\psi_3\rangle
\end{cases}
$$
If the result of the measurement is $\mathcal{M} = 1$, we know that our state is $|\psi_1\rangle$ since $\cos^2\left(\kappa_{12}\right) = 0$ and $\cos^2\left(\kappa_{13}\right) \rightarrow 0$
as $\alpha\rightarrow-\frac{\pi}{2}$.
Otherwise, we exclude $|\psi_1\rangle$ and apply the $\mathcal{PT}$-symmetric quantum state discrimination  for two quantum states, $|\psi_2\rangle$ and $|\psi_3\rangle$, which was previously developed ~\cite{bender2013pt}.

\section{Conclusions}

For the case of $N = 3$ pure quantum states, we are able to eliminate the state with the highest prior probability first by making the angle between two of the states zero in the limit when $\alpha\rightarrow-\frac{\pi}{2}$. This is an extension of the result~\cite{bender2007faster} when the two states $\begin{pmatrix}
1  \\
0
\end{pmatrix}
$
and 
$
\begin{pmatrix}
0  \\
1
\end{pmatrix}
$
have an angular separation of $\pi - 2|\alpha|$. After elimination of the state with the highest prior probability, the remaining two states can be distinguished by the two-state $\mathcal{PT}$-symmetric quantum state discrimination. In such a way, \textit{in principle,} we are able to find our state in at most 2 measurements and on average in $\left(2 - p\right)$ measurements with $p$ being the highest prior probability among the three states, $p = \underset{i \in [1, 3]}{max} \{p_i\}$. One has to keep in mind that some of the measurements may be inconclusive since the procedure involves the post-selection, so on practice more measurements may be required. Importantly, in our considerations we didn't make any assumptions on the geometry of our three states. Instead, we used an additional degree of freedom to adjust the geometry of these states before making measurements.

\end{document}